\documentclass{llncs}

\usepackage{graphicx}
\usepackage{caption,subfig}
\usepackage{amsmath}
\usepackage{mathptmx} % pretty math
\usepackage{cite} % pretty citation
\usepackage{color}
\usepackage{wrapfig}
\begin{document}

% start of the contributions
\mainmatter

%\title{DNA Computing with Reaction Dynamics}
\title{Reservoir Computing Approach to Robust Computation using Unreliable Nanoscale Networks}

\titlerunning{Reservoir Computing Approach to Robust Computation}

\author{Alireza Goudarzi\inst{1}, Matthew R. Lakin\inst{1}, \and Darko Stefanovic\inst{1,2}}

% abbreviated author list (for running head)
\authorrunning{Alireza Goudarzi et al.}

\institute{Department of Computer Science\\ University of New Mexico\\ \and Center for Biomedical Engineering\\ University of New Mexico\\ \email{alirezag@cs.unm.edu}}

\maketitle

% make caption smaller
\captionsetup{font=small}

%%%ABSTRACT
\begin{abstract}

As we approach the physical limits of CMOS technology,
advances in materials science and nanotechnology are making available a variety of unconventional computing substrates that can potentially replace top-down-designed silicon-based computing devices. Inherent stochasticity in the fabrication process and nanometer scale of these substrates inevitably lead to design variations, defects, faults, and noise in the resulting devices. A key challenge is how to harness such devices to perform robust computation.  We propose reservoir computing as a solution. In reservoir computing, computation takes place by translating the dynamics of an excited medium, called a reservoir, into a desired output. This approach eliminates the need for external control and redundancy, and the programming is done using a closed-form regression problem on the output, which also allows concurrent programming using a single device.  Using a theoretical model, we show that both regular and irregular reservoirs are intrinsically robust to structural noise as they perform  computation.

%\keywords{reservoir computing, molecular reservoir computing, DNA reservoir computing}

\end{abstract}

%%%%SECTION
\section{Introduction}

The approaching physical limits of silicon-based semiconductor technology are making conventional top-down designed  computer architecture prohibitive\cite{Haselman:2010p657}. Recent advances in materials science and nanotechnology suggest that unconventional computer architectures could be a viable technological and economical alternative. Some  proposed alternative architectures are based  on molecular switches and memristive crossbars \cite{0957-4484-14-4-311,Snider:2005qa} that possess  highly regular structure. Another emerging approach is self-assembly of nanowires and memristive networks  \cite{doi:10.1021/jp109207d,ADMA:ADMA201103053}, which results in irregular structure. Major obstacles to using such architectures are design variations, defects, faults, and susceptibility to environmental factors such as thermal noise and radiation \cite{itrs2011}. How should one {\em program} an unreliable system with unknown structure to perform reliable computation? Here we use a novel implementation of reservoir computing with sparse input and output connections to model self-assembled nanoscale systems and analyze their robustness to structural noise in the system.

Most approaches  assume knowledge of the underlying architecture and rely on   reconfiguration and redundancy to achieve programming and fault tolerance\cite{1230995,Oaloudek:2012fk,6144380,4617203}. There have been two recent proposals on how to program such devices to perform classification and logic operation using a ``black-box" approach\cite{Lawson:2006-04-01T00:00:00:1546-1955:272,6144633}. Both approaches are based on a theoretical model, called a randomly assembled computer (RAC), realized by a network of interacting nodes with sparse and irregular connectivity. All nodes are initialized to zero and update their state according to a global clock, and each node calculates its next state using its transfer function and connections to other nodes. Three types of external signals are connected to randomly chosen nodes: inputs, outputs, and controls. The task is to program the device to compute the desired output for a given input using a proper control signal. The optimal control signal will  modify the propagation of input across the network so that input is processed as required and the desired result is presented at the output. The optimal control signals are computed using simulated annealing. The key property of this model is sparse random external interfaces, i.e., input, output, and controls. The model's only fundamental and reasonable assumption is that there is enough connectivity that the input and control signals can propagate through the network and reach the output. This model has shown impressive performance and inherent robustness to noise\cite{Lawson:2006-04-01T00:00:00:1546-1955:272}. In RAC, the computation takes place by initializing the network with a fixed state and presenting the input signal to the network, and then the network runs until the output is produced. This cycle is repeated for each new input pattern. The computation is therefore sensitive to the initial state of the network and the control signals must be calculated based on the desired computation, the structure, and the initial state of the network.

We propose the reservoir computing (RC) paradigm\cite{springerlink:10.1007} as an alternative programming approach to unconventional and irregular architectures. RC lets the network dynamics be perturbed by the input signal and maps the network states to the desired output using closed-form linear regression. In addition to the connectedness assumption from RAC, we require  the network to have a slowly converging dynamics. RC provides several advantages over RAC. In RC, the computation is not sensitive to the initial state of the system and there is no need for control signals, which leads to  simpler design and implementation. Also, the training is done in a closed-form regression and does not need an iterative process. Moreover, nonlinear computation is inherently enabled by the network dynamics acting as a recursive kernel and extracting nonlinear features of the input signal\cite{Hermans:2011fk}. Noise in the input, the network states, and the interactions between the nodes can be treated using a regularization term and can be scaled to achieve the best performance.  This is particularly attractive, because RC depends on the dynamics to compute, and structural change may have adverse effects on the dynamical regime of the system, which would normally require retraining the network. In addition, the programming is performed on the output instead of the task-specific control of the network, and therefore we can compute multiple functions simultaneously using the same device. In contrast  to existing RC implementations\cite{Lukosevicius:2009p1443,verstraeten2007}, the novelty of our work  is the consideration of sparse input and output to model unconventional computer architectures, and the analysis of robustness in the presence of structural noise in the network, possibly due to thermal noise and radiation that change the electrical properties of the network. In classical implementations of RC, the input and output are connected to all the internal nodes and the system is assumed to operate in a noise-free environment. We demonstrate the performance and robustness of RC using regular and irregular networks and analyze the memory capacity and nonlinear computational performance of the system subject to structural noise. Our results show that RC can be a viable approach to using self-assembled and nanoscale substrates to implement  robust, special-purpose signal processing devices.
%%%
%\begin{wrapfigure}{R}{0.42\textwidth}
\begin{figure}[ht]
\centering
\includegraphics[width=.8\textwidth]{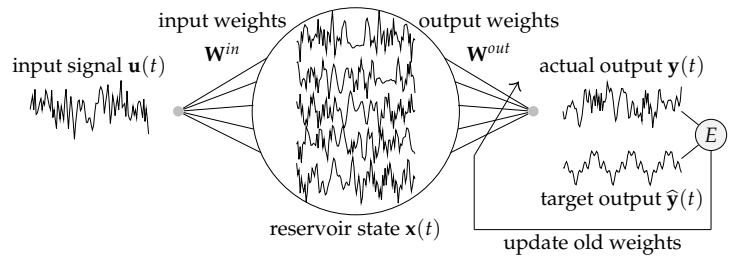}

\caption{Computation in a reservoir computer. The reservoir is an excitable dynamical system with $N$ readable output states represented by the vector ${\bf X}(t)$. The input signal ${\bf u}(t)$ is fed
into one or more points $i$ in the reservoir with a corresponding weight $w^{in}_i$
denoted with weight column vector ${\bf W}^{in}=[w^{in}_i]$.}
\label{fig:fig1}
\end{figure}
%%%

\section{Background}
\label{sec:background}

Reservoir computing was independently introduced by Maass, Natschl{\"a}ger, and Markram\cite{Maass:2002p1444} and by Jaeger\cite{Jaeger:2001p1442}. Echo state networks (ESN) are one of the most popular RC paradigms, and have shown promising results in time series computing and prediction
\cite{Wyffels20101958,Jaeger02042004}, voice recognition\cite{Paquot:2012fk}, nonlinear system identification\cite{Jaeger:2003p1447}, and robot control\cite{Dasgupta:2012fk}. An ESN\cite{Jaeger:2003p1447,Jaeger:2001p1446,Jaeger:2002p1445,verstraeten2007} consists of an input-driven recurrent neural network, which acts as the reservoir, and a readout layer that reads the reservoir states and produces the output. Unlike a classical recurrent neural network, where all the nodes are interconnected and their weights are determined during a training process, in an ESN the nodes are interconnected using random weights and random sparse connectivity between the nodes. The input and reservoir connections are initialized and fixed, usually with no further adaptation.

%%%%%%%
\begin{figure}[t]%{R}{0.5\textwidth}
\centering
\includegraphics[width=0.5\textwidth]{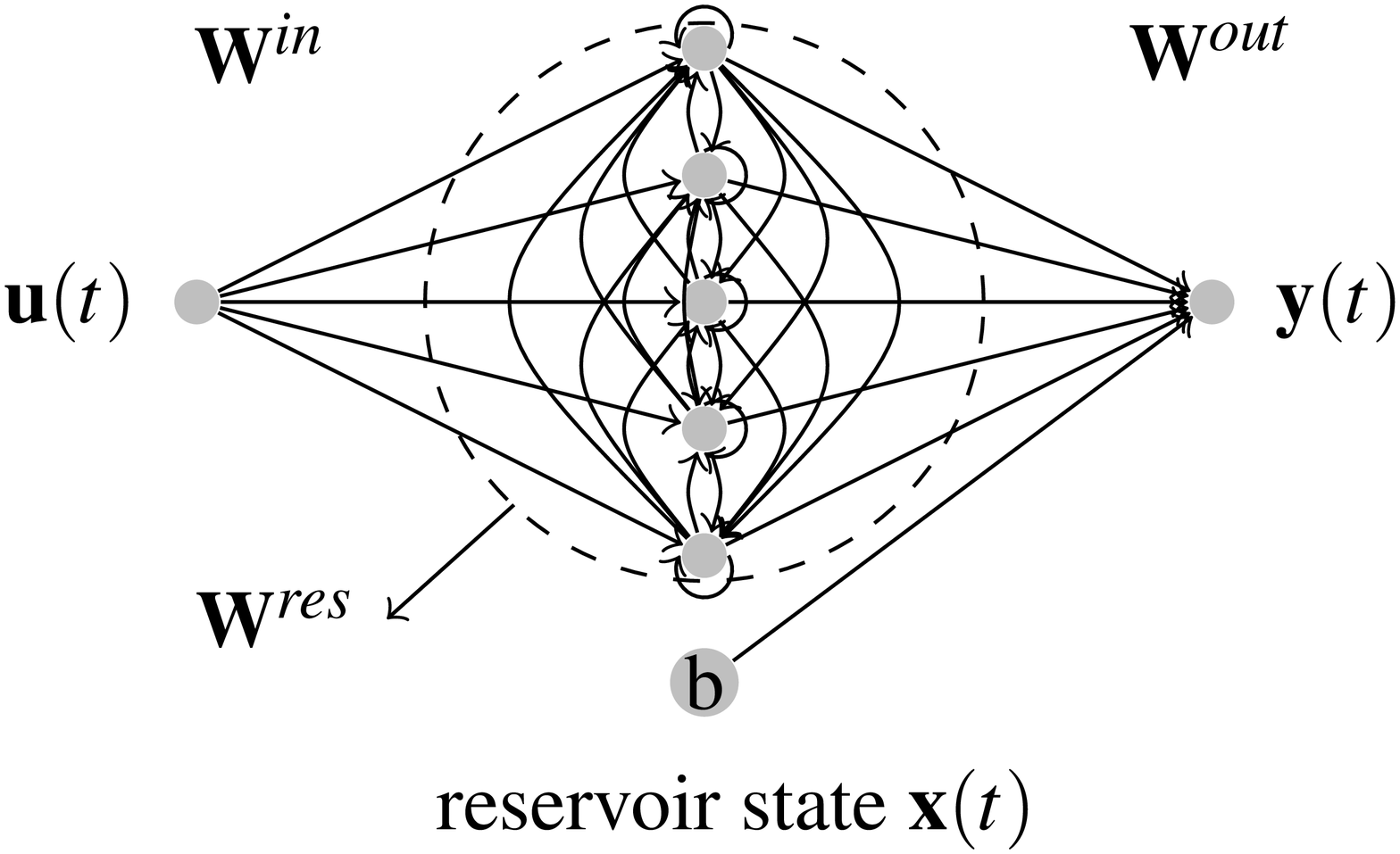}
\caption{Schematic of an ESN. A dynamical core called a reservoir is driven by input signal ${\bf u}(t)$. The states of the reservoir ${\bf x}(t)$ extended by a constant $1$ and combined linearly to produce the output ${\bf y}(t)$. The reservoir consists of $N$ nodes interconnected with a random weight matrix ${\bf W}^{res}$. The connectivity between the input and the reservoir nodes is represented with a randomly generated weight matrix ${\bf W}^{in}$. The reservoir states and the constant are connected to the readout layer using the weight matrix ${\bf W}^{out}$. The reservoir and the input weights are fixed after initialization, while the output weights are learned using a regression technique.}
\label{fig:ESN}
\end{figure}
%%%%%%

Figure~\ref{fig:ESN} shows a schematic of an ESN. The readout layer is usually a linear combination of the reservoir states. The readout weights are determined using supervised learning techniques, where the network is driven by a teacher input and its output is compared with a corresponding teacher output to estimate the error. Then, the weights can be calculated using any closed-form regression technique\cite{Jaeger:2002p1445} in offline training contexts, or using adaptive techniques if online training is needed\cite{Jaeger:2003p1447}.  Mathematically, the input-driven reservoir is defined as follows. Let $N$ be the size of the reservoir. We represent the time-dependent inputs as a column vector ${\bf u}(t)$, the reservoir state as a column vector ${\bf x}(t)$, and the output as a column vector ${\bf y}(t)$. The input connectivity is represented by the matrix ${\bf W}^{in}$ and the reservoir connectivity is represented by an $N\times N$ weight matrix ${\bf W}^{res}$. For simplicity, we assume  one input signal and one output, but the notation can be extended to multiple inputs and outputs. The time evolution of the reservoir is given by:

\begin{equation}
{\bf x}(t+1) = f({\bf W}^{res}\cdot  {\bf x}(t) + {\bf W}^{in}\cdot {\bf u}(t)),
\end{equation}
where $f$ is the transfer function of the reservoir nodes that is applied element-wise to its operand. This is usually the hyperbolic tangent, but sigmoidal or linear functions can be used instead. The output is generated by the multiplication of  an output weight matrix ${\bf W}^{out}$ of length  $N+1$ and the reservoir state vector $x(t)$ extended by a constant $1$ represented by ${\bf x}'(t)$:
\begin{equation}
{\bf y}(t) = {\bf W}^{out}\cdot {\bf x}'(t).
\label{eq:output}
\end{equation}

The output weights ${\bf W}^{out}$ must be trained using a teacher input-output pair using regression\cite{verstraeten2007,PSP:2043984,5629375}. This regression can be performed in closed form and therefore ESN training is very efficient compared with classical recurrent neural network training, which requires a time-consuming iterative process\cite{846741}.

In ESN, the reservoir acts as a recursive kernel which creates an expressive spatiotemporal code for the input signal\cite{Hermans:2011fk}. In ESNs, to create the required spatiotemporal feature space, the reservoir must enjoy the so-called echo state property\cite{Jaeger:2001p1446} (ESP): over time the asymptotic state of the reservoir  depends only on the history of the input signal ${\bf u}(t)$, i.e., the dynamics is independent of the initial state of the network. Jaeger\cite{Jaeger:2001p1446} showed that to satisfy this condition, the reservoir weight matrix ${\bf W}^{res}$ must have the spectral radius $\lambda^{max}<1$ and the largest singular values $\sigma^{max}<1$.

\clearpage
\section{Experimental Setup}
\label{sec:setup}

\subsection{Reservoir Generation and Inducing Noise}
\begin{wrapfigure}{R}{0.42\textwidth}
\centering
\subfloat[]{
\includegraphics[width=1in]{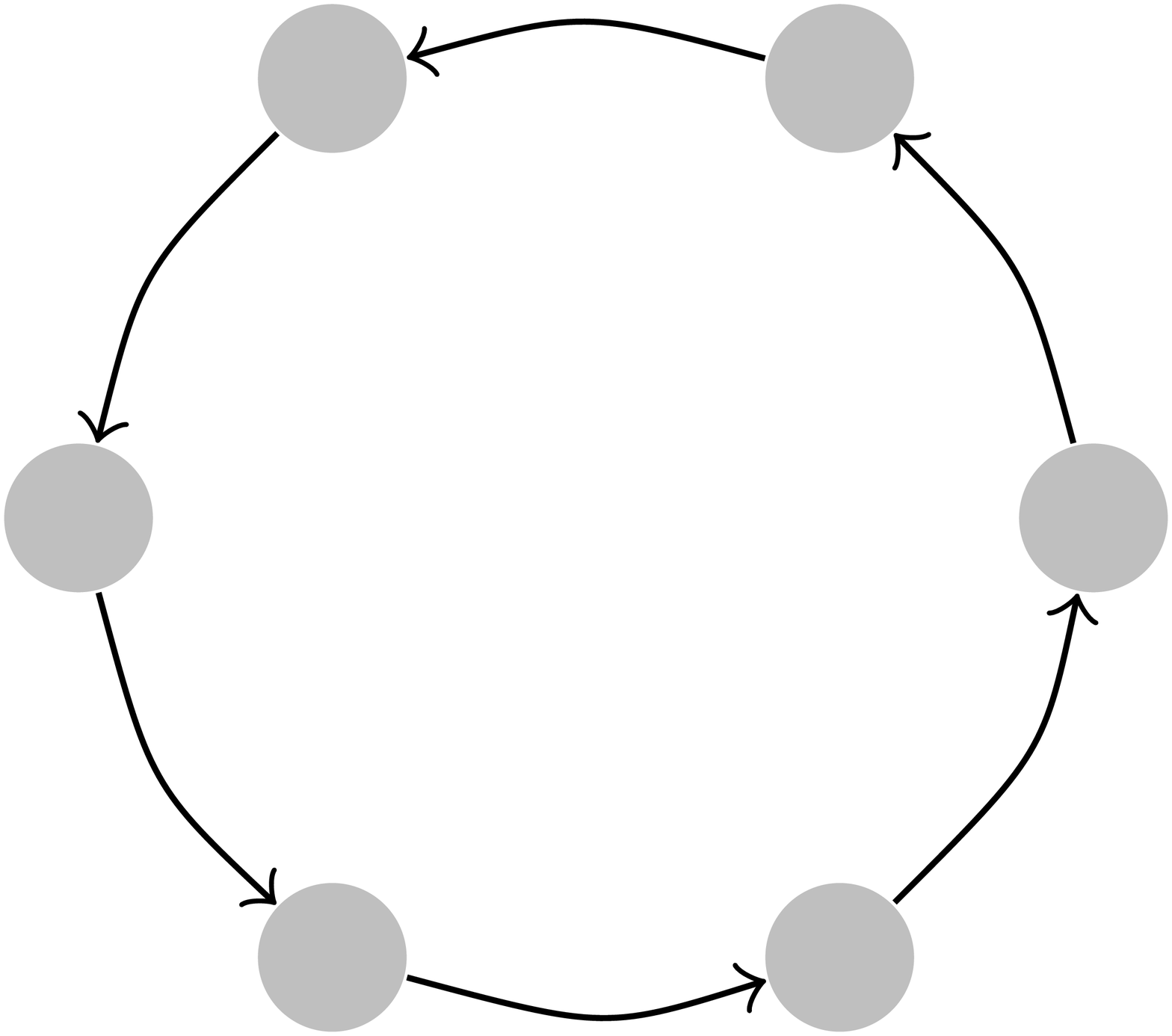}
}
\subfloat[]{
\includegraphics[width=1in]{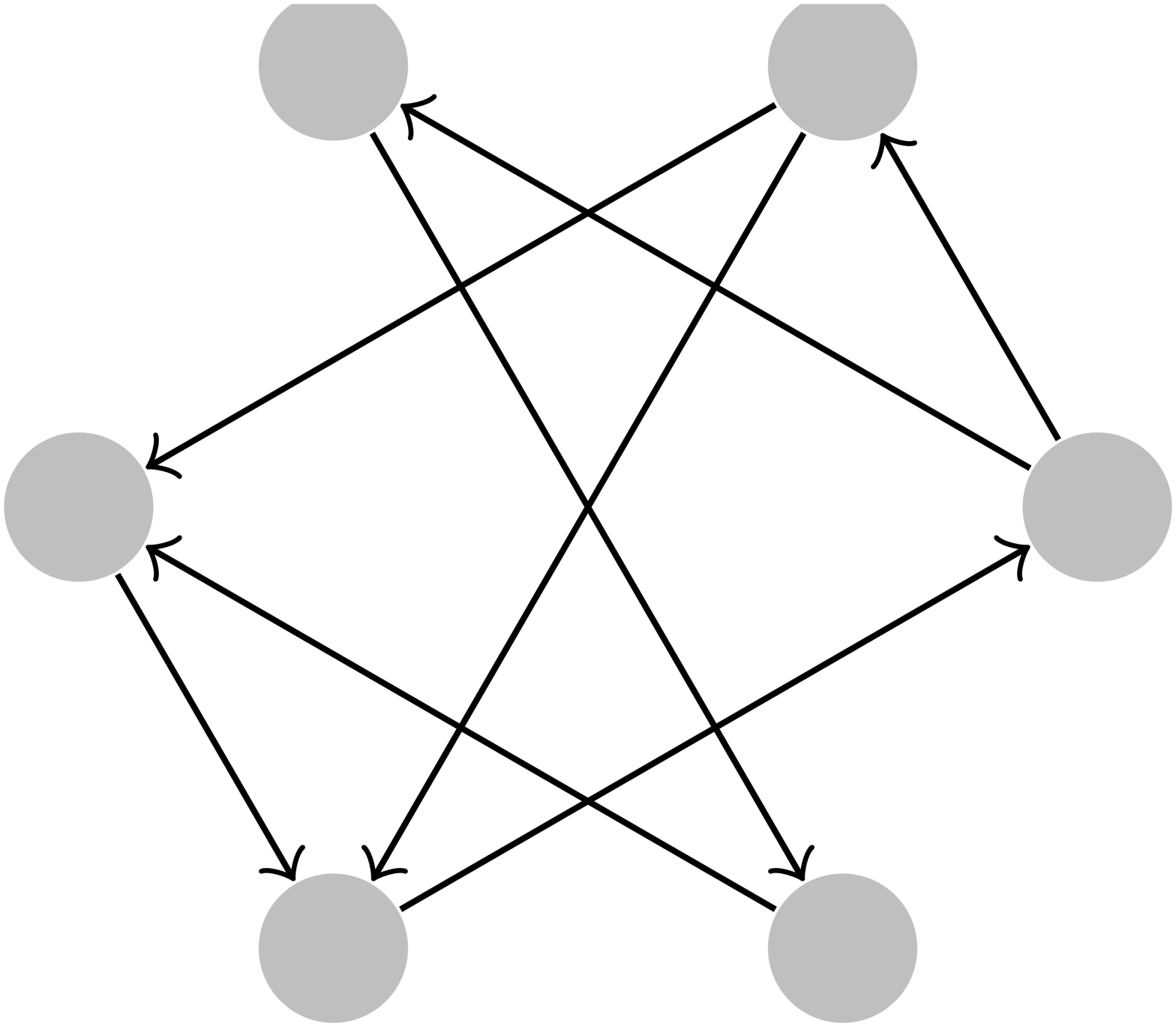}
}
\caption{Schematic for structure of a SCR reservoir (a) and a random sparse reservoir (b).}
\label{fig:conn}
\end{wrapfigure}
Similar to \cite{5629375}, we use RC  with a simple cycle reservoir (SCR) and ESNs with sparse randomly generated reservoirs for our experiments. We specify the number of reservoir nodes by $N$. We use the hyperbolic tangent  transfer function in both models. In SCR, the reservoir nodes are connected in a cycle and have identical weights $r$, $0<r<1$. It has been shown\cite{5629375} that despite the simplicity of this model its performance is comparable to sparse random reservoirs. In an ESN, a fraction $l$ of all possible connections are chosen to have non-zero weights and the rest of the connections are assigned zero; half of the non-zero weights are assigned $-0.47$ and the other half are assigned $+0.47$. The choice of $\pm0.47$ ensures ESP, which otherwise can be achieved by scaling the reservoir weight matrix as ${\bf W}^{res}\leftarrow\lambda{\bf W}^{res}/\lambda^{max}$ , where $\lambda^{max}$ is the spectral radius of  ${\bf W}^{res}$ and $\lambda$ is the desired spectral radius $0<\lambda<1$. The non-zero weights are chosen independently and randomly, which results in an Erd{\"o}s-R{\'e}nyi network structure\cite{Erdos:1959p1849}. Figure~\ref{fig:conn} illustrates the structure of SCR and random sparse reservoirs. For both models, the input signal is connected to half of the nodes that are picked randomly and the input weights are chosen from the set $\{-v,+v\}$ according to Bernoulli distribution, where $v$ is the input coefficient. For our experiments, we use  sets of input coefficients $V$, SCR reservoir weights $R$, and ESN spectral radii $\Lambda$ varying in the range $[0.1,0.9]$ with $0.1$ increments.

%\subsection{Inducing Structural Noise}
To study the effect of structural noise on RC performance, we add a white noise term, with standard deviation $\sigma$, to $n$ randomly chosen non-zero entries of ${\bf W}^{res}$ at each time step $t$. This will cause the non-zero entries of ${\bf W}^{res}$ to vary around their initial value according to a normal distribution. Our motivation for this is to model short term temporal variations in the structural properties of nanoscale networks. These variations are known to follow a normal distribution\cite{4447311}.  We choose $n$ for each experiment to make sure the fraction of noisy weights  is constant across all reservoirs.

\subsection{Simulation, Training, and Evaluation}

To evaluate the performance of each model, we generate $50$ streams of random numbers picked uniformly from the interval $[-1,+1]$. For each stream a new ESN or SCR was instantiated with randomized states uniformly picked from the interval $[-1,+1]$. The system was then driven for $T+2,000$ time steps. The first $T$ steps were then discarded to account for the transient period, where $T$ is chosen to be half of the reservoir size $N$. We randomly chose half the reservoir nodes to read reservoir states; the states of these nodes were then collected and augmented with a constant $1$ as inductive bias and arranged row-wise into a matrix ${\bf X}$, which was used for calculating the output weights ${\bf W}^{out}$ given by:
\begin{equation}
{\bf W}^{out} = {\bf M}\cdot \widehat{\bf y} ,
\end{equation}
where $\widehat{\bf y}$ is the expected output. The matrix ${\bf M}$ is usually calculated using either an ordinary linear regression technique as $({\bf X}^T\cdot{\bf X})^{-1}\cdot{\bf X}^T$, or a ridge regression technique as $({\bf X}^T\cdot{\bf X} + \gamma^2{\bf I})^{-1}\cdot{\bf X}^T$, where  $\gamma$ is the regularization factor, and ${\bf I}$ is the identity matrix of order $N+1$. In general the spectra of ${\bf X}^T{\bf X}$ should be studied to choose an appropriate inversion technique. We found that using the Penrose-Moore pseudo-inverse of ${\bf X}$ for ${\bf M}$, which minimizes its norm, produces the most stable results. We calculated this using MATLAB's {\em pinv} function. To test performance, we drove the system for another $T+2000$ time steps of each stream,  created the matrix ${\bf X}$ as before, and calculated the output as in Equation~\ref{eq:output}. We evaluate the robustness of SCR and ESN as percent change in their performance for two different tasks described below.

\subsubsection{Memory Capacity (MC).}
Jaeger\cite{Jaeger:2001p1446} defined the memory capacity task to quantify the short-term memory of the reservoir in ESN by measuring how well the network can reconstruct the input after $\tau$ number of time steps. The coefficient of determination between the input and a $\tau$-delayed version of the input as   output of ESN is:
\begin{equation}
MC_{\tau} = \frac{\text{cov}^2(u(t-\tau),y(t))}{\text{var}(u(t))\text{var}(y(t))}.
\end{equation}
 The total memory capacity of a network is then given by:
\begin{equation}
MC = \sum_{\tau=1}^\infty MC_{\tau}.
\end{equation}
Assuming a zero-centered uniformly random stream as input, the memory capacity for ESN is bounded by the size of the reservoir $MC<N$\cite{Jaeger:2001p1446} , and $N-1<MC<N$ for SCR\cite{5629375}. However, the empirical  values vary based on experimental conditions. We derive the networks with the input streams as described previously in this section and we measure the MC for both ESN and SCR of size $N=50$, using a finite sum of $MC_\tau$ up to $\tau=200$.  We can then measure memory robustness as the ratio of memory capacity of noisy systems $MC$ to the noise-free  systems $MC^*$ for ESN and SCR as follows:
\begin{equation}
\Gamma_{MC}^{ESN}(v,\lambda) =  \frac{MC(v,\lambda)}{MC^{*}(v,\lambda)} \text{ and } \Gamma_{MC}^{SCR}(v,r) =  \frac{MC(v,r)}{MC^{*}(v,r)}
\end{equation}
where $k$ is the fraction of noise-induced connections and $\sigma$ is the standard deviation of the noise.  We let $MC(v,\lambda)$ and $MC(v,r)$ denote the memory capacity of ESN with parameters $v$ and $\lambda$, and memory capacity of SCR with parameters $v$ and $r$, respectively.

%\begin{wrapfigure}{R}{0.42\textwidth}
%\begin{figure}[t]
%\centering
%\includegraphics[width=0.6\textwidth]{rawmc}
%\caption{Delayed-input reconstruction capacity $MC_\tau$ of SCR and ESN in noisy and noise-free conditions. In all cases the $MC_\tau$ shows a sharp drop after a critical $\tau^*$ to a value smaller then $0.1$. To compute the total memory capacity $MC$, we consider all $MC_\tau<0.1$ as zero.}
%\label{fig:rawmc}
%\end{figure}

\subsubsection{Nonlinear Autoregressive Moving Average (NARMA).}

NARMA is a nonlinear task with long time lag designed to measure neural network capability to compute nonlinear functions of previous inputs and outputs. The 10-th order NARMA system NARMA10 is defined as follows:
\begin{equation}
y(t)=0.3 y(t-1)+0.05 y(t-1) \sum_{i=1}^{10}y(t-i)+1.5 u(t-10)u(t-1)+0.1.
\label{eq:narma}
\end{equation} The input $u_t$ is
drawn from a uniform distribution in the interval $[0,0.5]$. To generate the input for this task, we shift our input streams by $2$ and divide them by $4$ to ensure the values are in the internal $[0,0.5]$. We calculate the performance of ESN and SCR on this task by the test error  measured by the normalized mean squared error (NMSE) given by:
\begin{equation}
NMSE = \frac{\left\langle (y(t) - \widehat{y}(t))^2 \right\rangle}{
\text{var}(\widehat{y}(t))}.
\end{equation}
If the mean squared error of the output is larger than the variance of the target output then  $NMSE>1$, in which case we consider the $NMSE=1$ to simplify our analysis. Once again we measure robustness with respect to the error as the ratio between the error of a noisy system $NMSE$ to the error of a noise free-system $NMSE^*$ as follows:
\begin{equation}
\Gamma_{NMSE}^{ESN}(v,\lambda) =  \frac{NMSE^*(v,\lambda)}{NMSE^(v,\lambda)} \text{ and } \Gamma_{NMSE}^{SCR}(v,r) =  \frac{NMSE^*(v,r)}{NMSE(v,r)},
\end{equation}
using  $NMSE(v,\lambda)$ and $NMSE(v,r)$ as shorthand for the performance of ESN with parameters $v$ and $\lambda$, and the performance of SCR with parameters $v$ and $r$, respectively.

\section{Results}
\label{sec:result}

\begin{figure}[ht]
\centering
\def \w {1.7in}
\subfloat[]{
\includegraphics[width=\w]{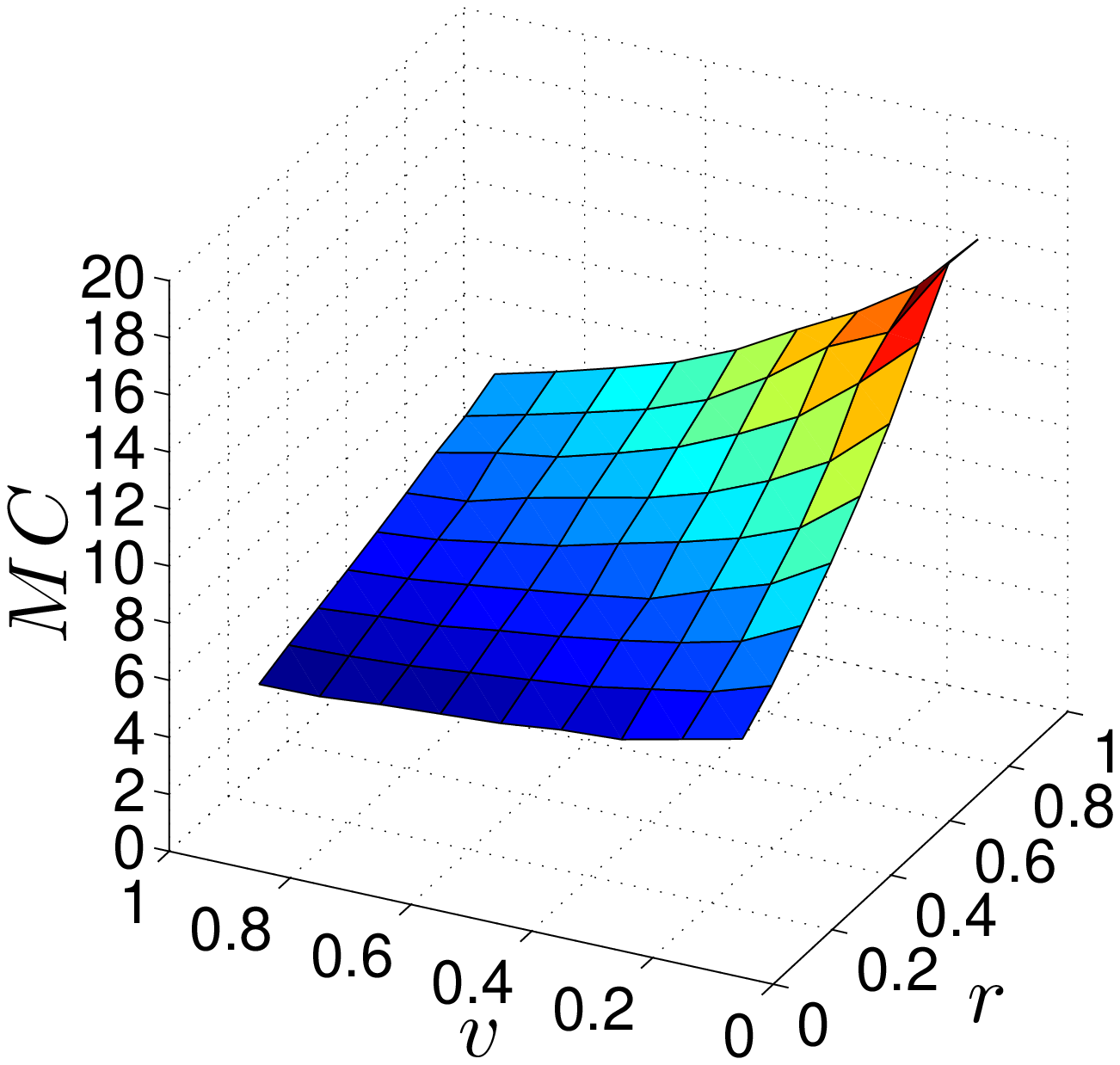}
\label{fig:scrmcnonoise}
}
\subfloat[]{
\includegraphics[width=\w]{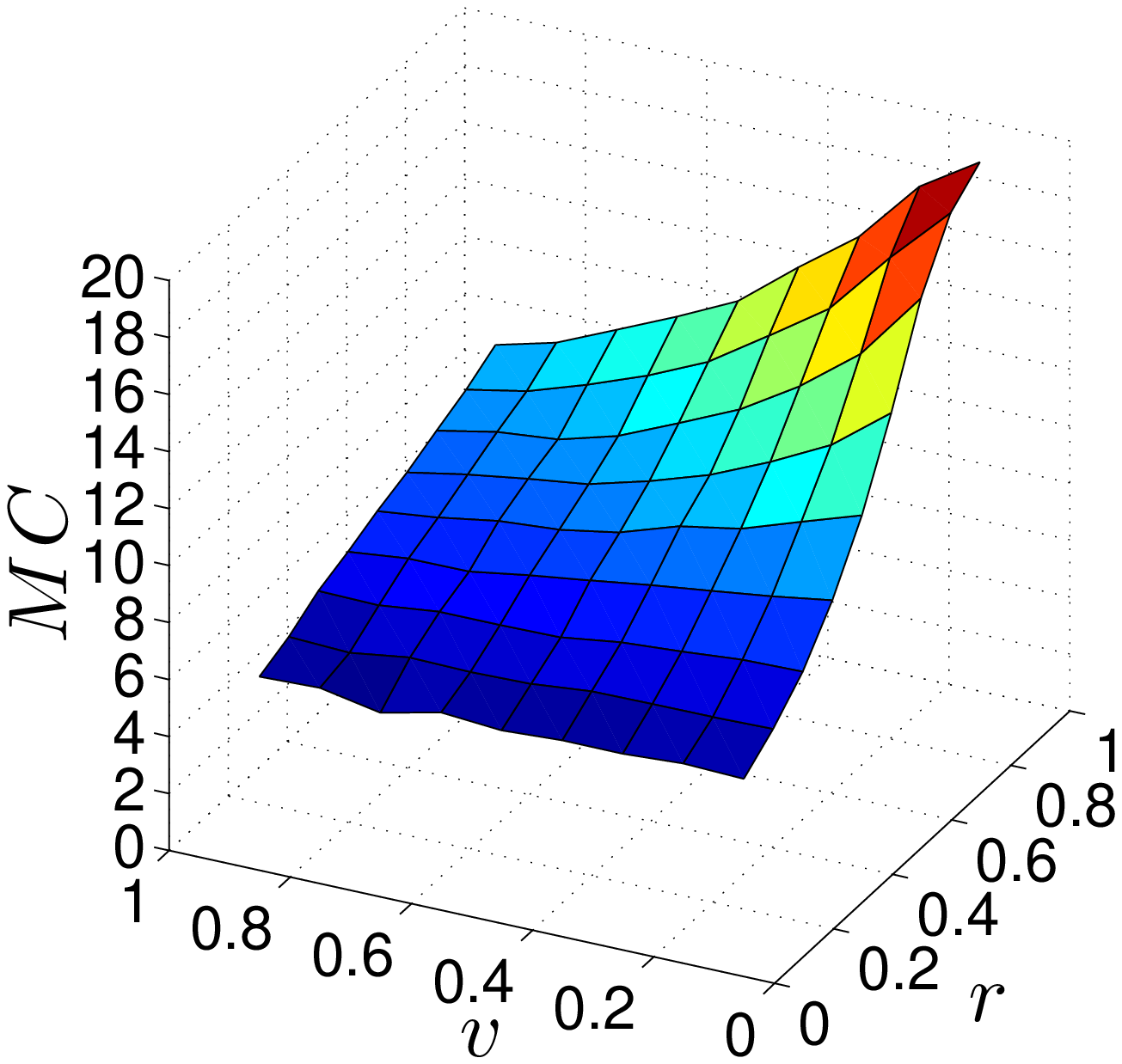}
\label{fig:scrmcnoisy}
}
\subfloat[]{
\includegraphics[width=\w]{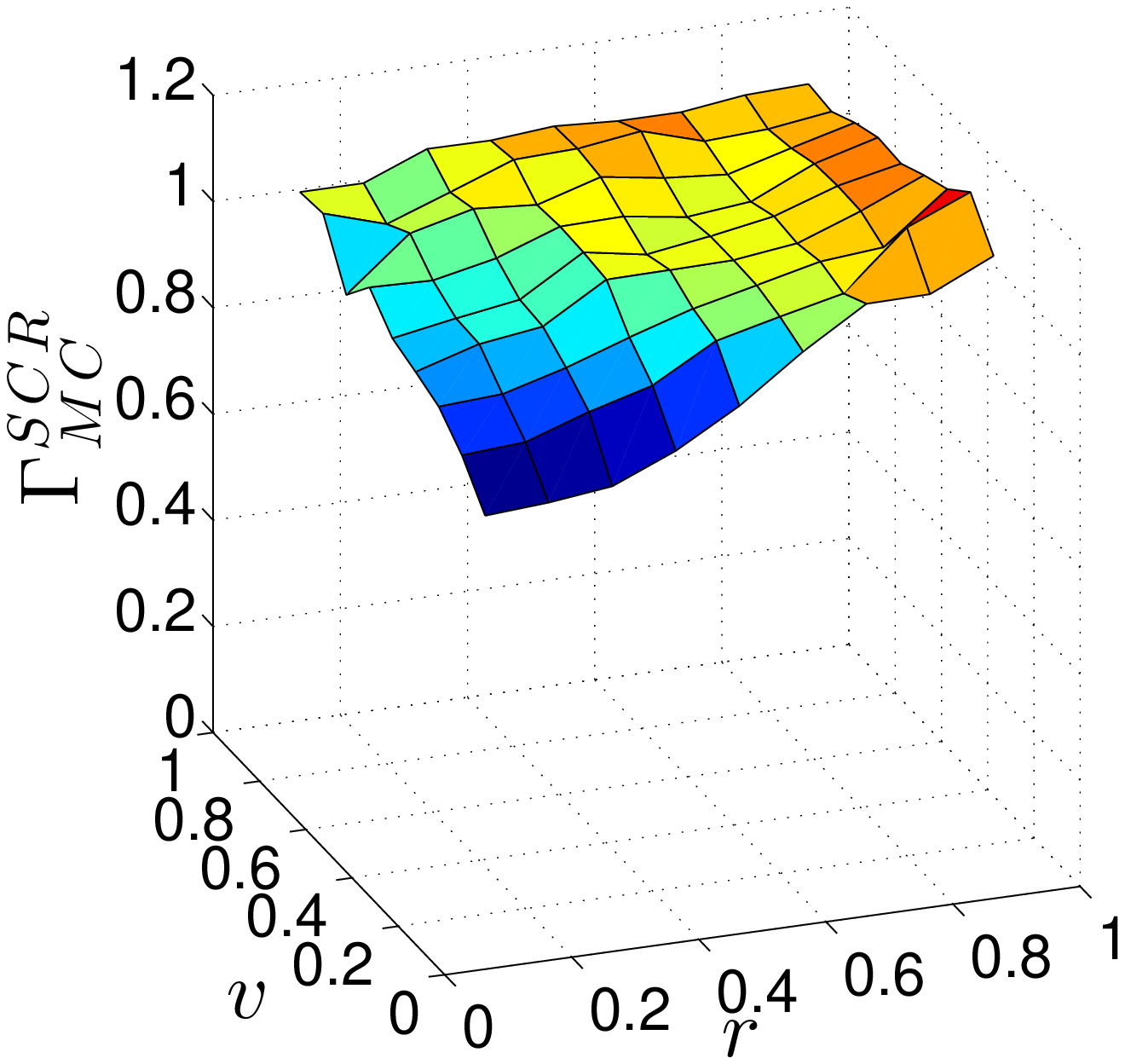}
\label{fig:scrmcratio}
}\\
\subfloat[]{
\includegraphics[width=\w]{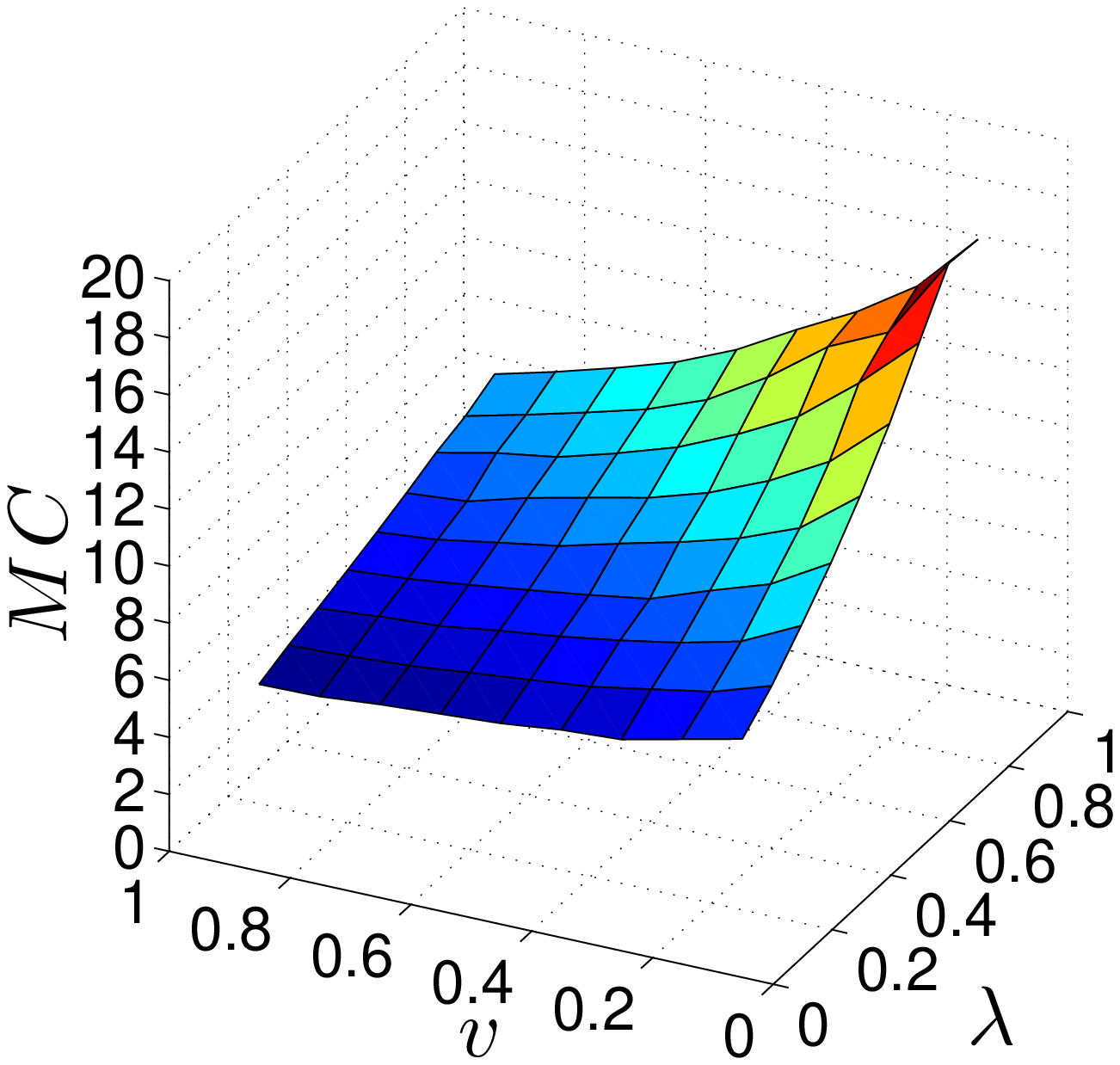}
\label{fig:esnmcnonoise}
}
\subfloat[]{
\includegraphics[width=\w]{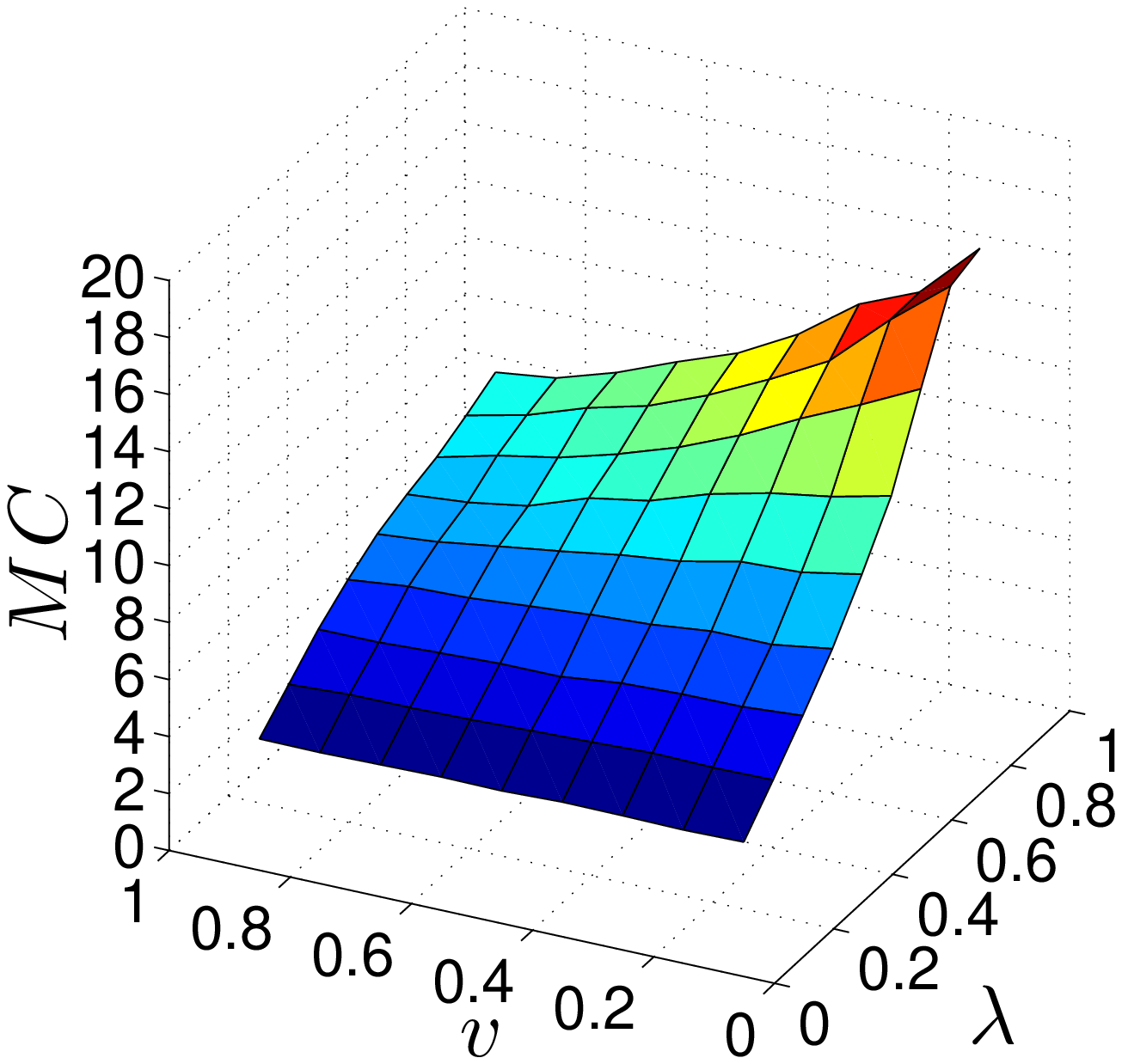}
\label{fig:esnmcnoisy}
}
\subfloat[]{
\includegraphics[width=\w]{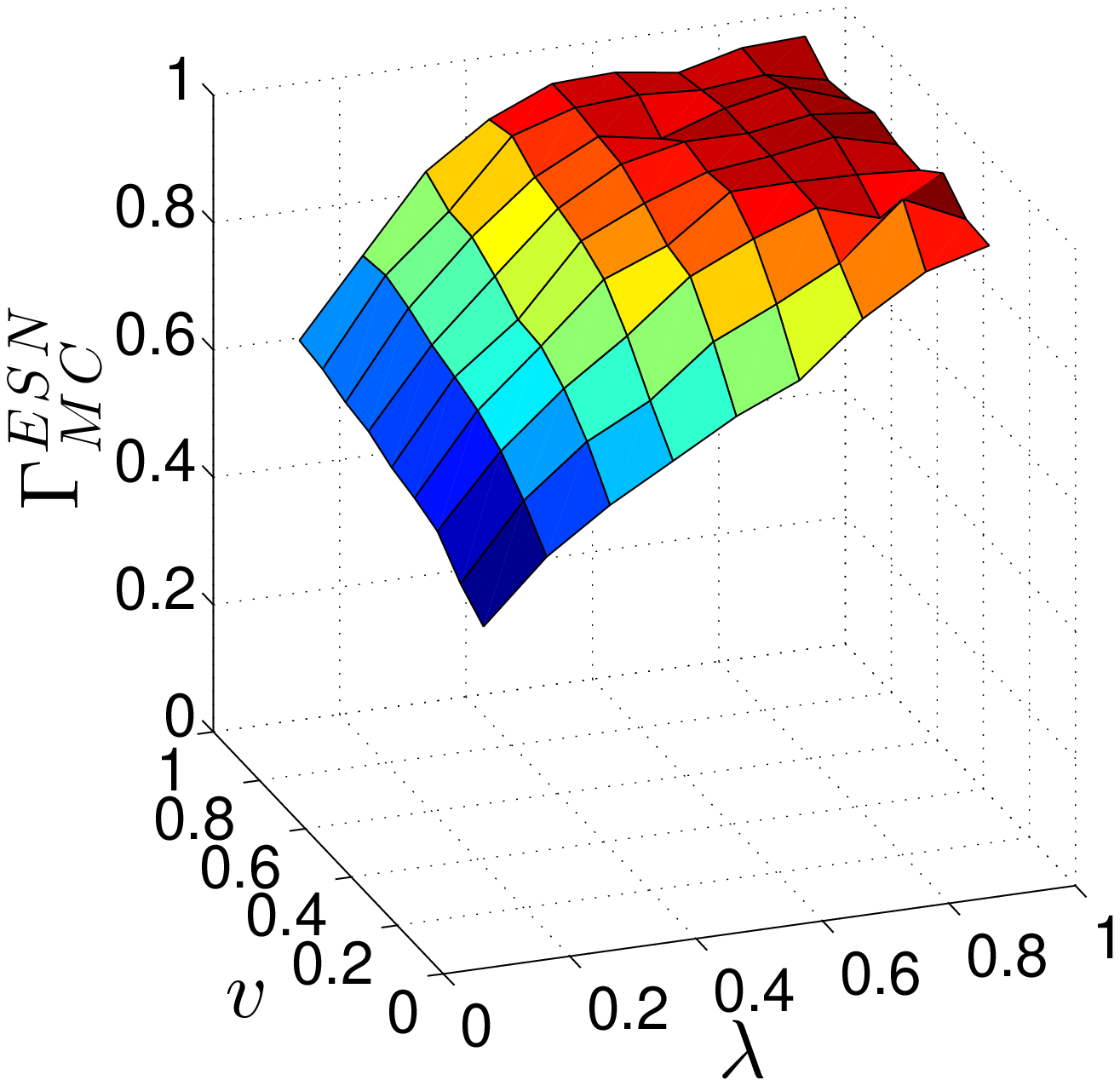}
\label{fig:esnmcratio}
}
\caption{Memory capacity (MC) in noise-free  SCR (a) and SCR with structural noise (b). The ratio $\Gamma_{MC}^{SCR}(v,r)$ showing the overall variation of the $MC$ between noisy and noise-free  conditions (c). MC in noise-free  ESN (d) and ESN with structural noise (e). The ratio $\Gamma_{MC}^{ESN}(v,\lambda)$ showing the overall variation of the $MC$ between noisy and noise-free  conditions (f). SCR memory is robust to noise for $r>0.6$ and $v\ge 0.1$ and ESN memory is robust to noise for $\lambda>0.8$ and $v\ge 0.1$.}
\label{fig:mc}
\end{figure}

First we analyze the memory capacity in SCR and ESN under  structural noise. All the results in this section are the average value over $60$ runs as described in Section~\ref{sec:setup}. Figure~\ref{fig:scrmcnonoise} shows the memory capacity of SCR for reservoirs of size $N=50$ without any structural noise. The MC shows a nonlinear increase for increasing $r$ and decreasing $v$ up to $r=0.8$ and $v=0.1$, where the MC reaches its maximum $MC=17.15$. Figure~\ref{fig:scrmcnoisy} shows the memory capacity of SCR under noisy conditions where at each time step a single randomly chosen node is perturbed with a white noise with standard deviation $\sigma=0.01$.  For suboptimal $r$ and $v$, the noise distorts the memory of the system, resulting in lower memory capacity, whereas for optimal parameters, the memory capacity increases due to the regularization effect of noise terms on the regression; in fact, at its peak memory capacity is $MC=19.74$. According to the ratio $\Gamma_{MC}^{SCR}$, shown in Figure~\ref{fig:scrmcratio}, for $r>0.6$ and $v\ge0.1$ the noise improves the memory capacity.
Figure~\ref{fig:esnmcnonoise} shows the memory capacity of ESNs of size $N=50$ and connection fraction $l=0.2$. Due to the variation inside the reservoir, the memory capacity surface for ESNs is not as smooth as the MC surface for SCRs. In ESNs, the memory capacity increases nonlinearly with increasing $\lambda$ and decreasing $v$ and reaches its maximum $MC=17.15$ at $\lambda=0.8$ and $v=0.1$. Figure~\ref{fig:esnmcnoisy} shows the memory capacity of ESNs under noisy conditions. At each step $n=10$ connections are perturbed using a white noise of standard deviation $\sigma=0.01$ to achieve the same noise level as for SCR. The effect of noise in ESN is slightly higher. The memory capacity changes slowly from $MC=3.40$ to its maximum $MC=16.72$ for $\lambda=0.9$. According to Figure~\ref{fig:esnmcratio} for all $v$ and $\lambda>0.8$ the MC is not decreased significantly in noisy conditions. In summary, both SCR and ESN  are highly robust to structural noise.

\begin{figure}[t]
\centering
\def \w {1.7in}
\subfloat[]{
\includegraphics[width=\w]{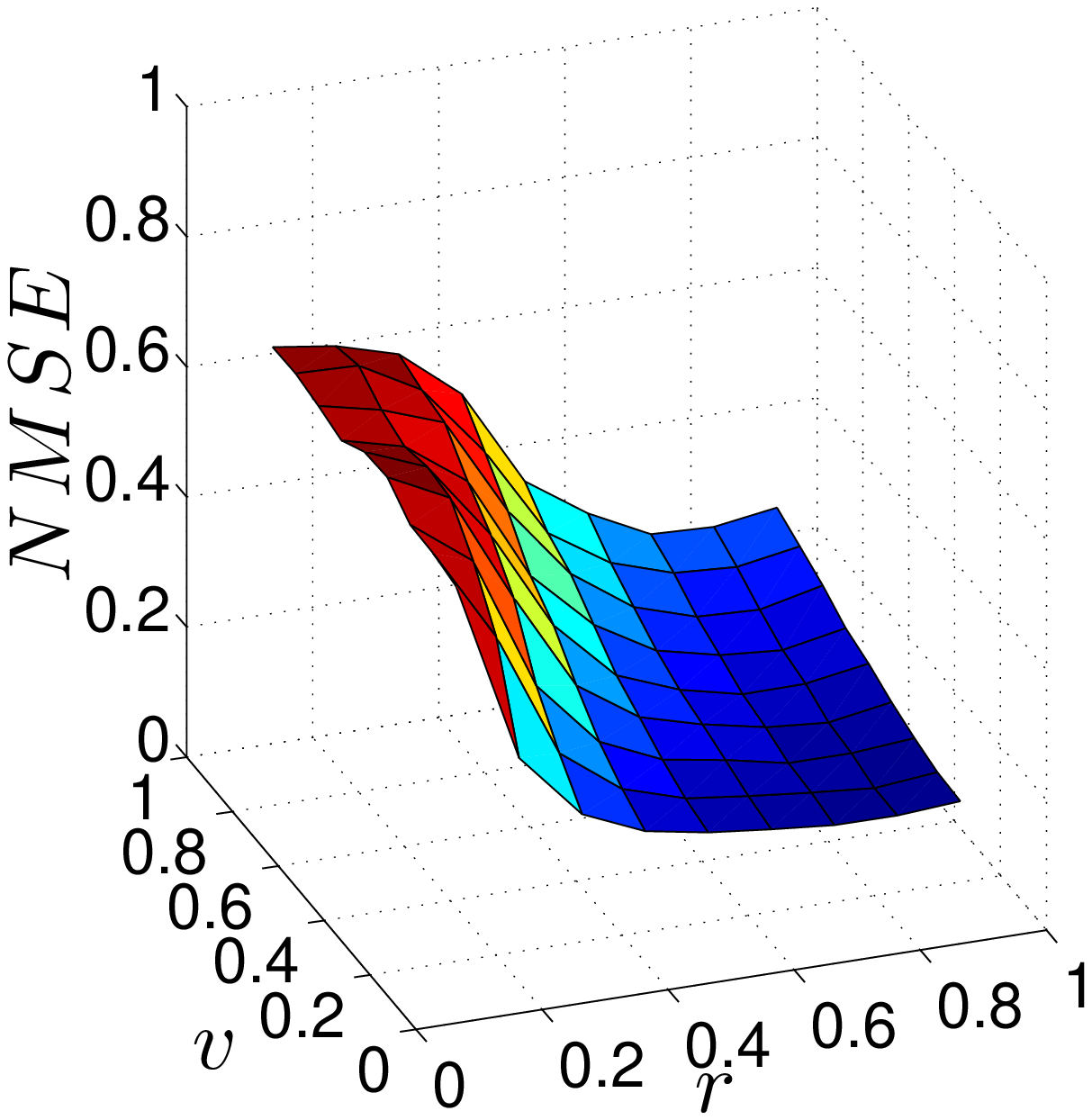}
\label{fig:scrnarmanonoise}
}
\subfloat[]{
\includegraphics[width=\w]{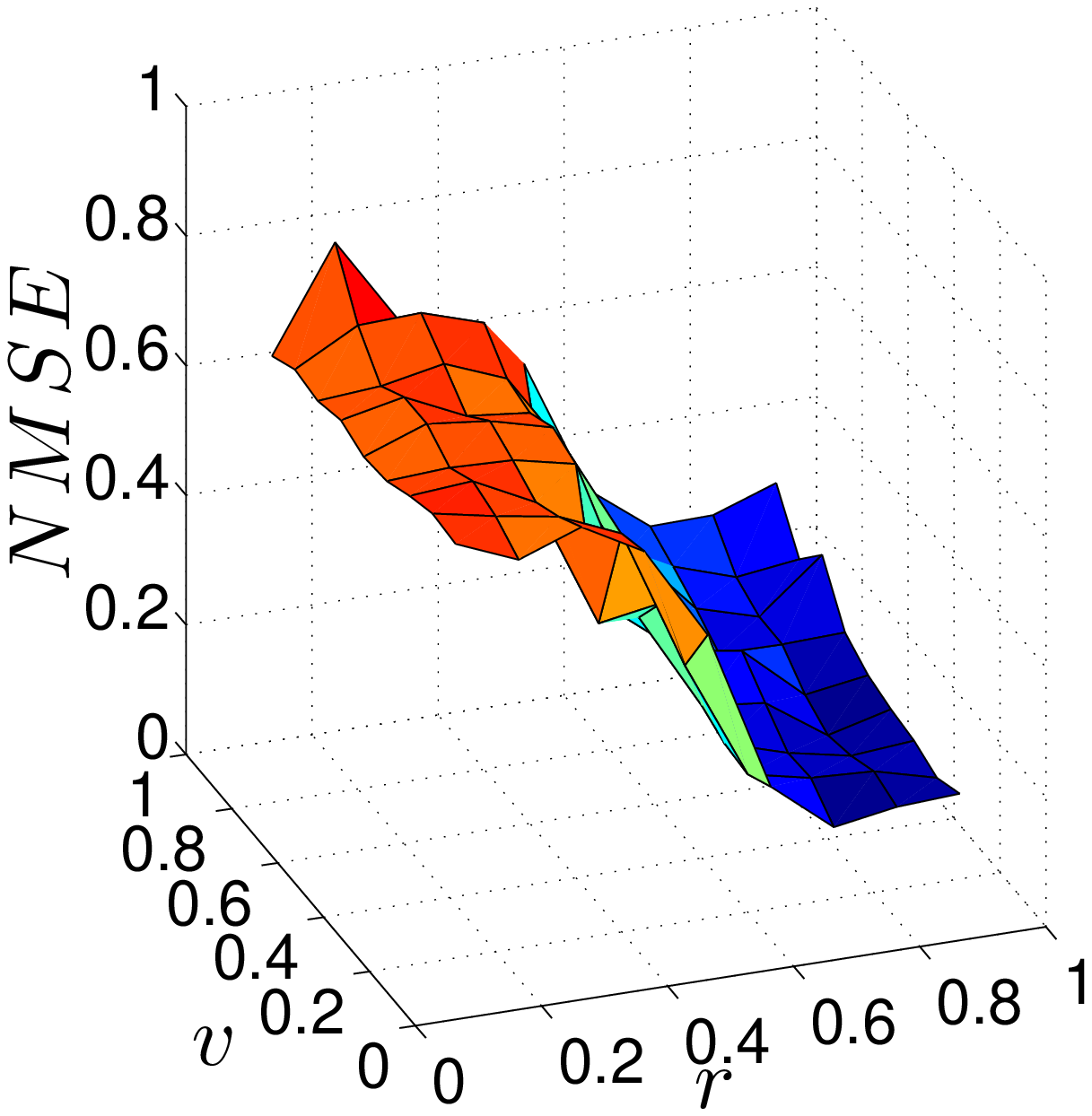}
\label{fig:scrnarmanoisy}
}
\subfloat[]{
\includegraphics[width=\w]{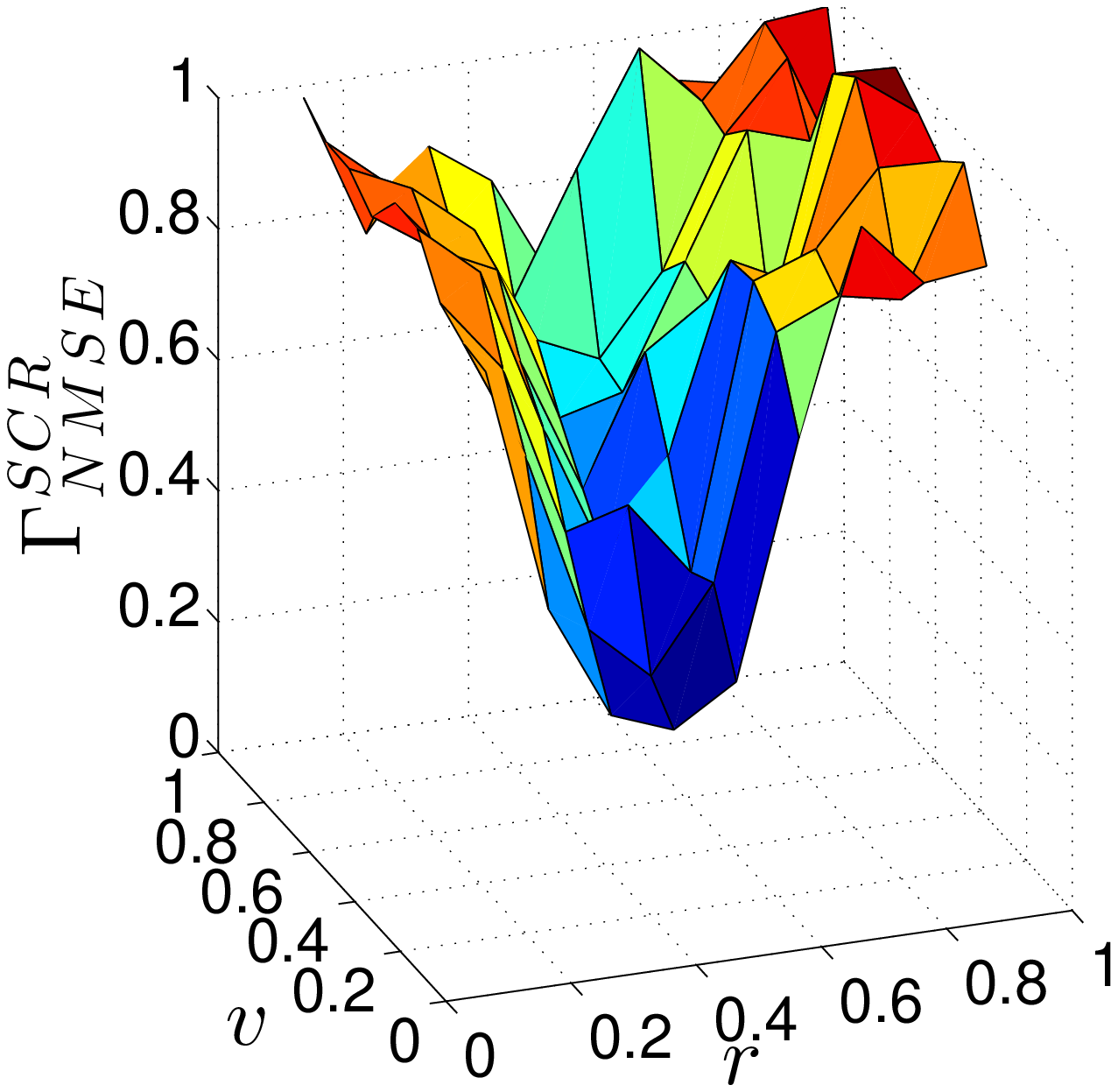}
\label{fig:scrnarmaratio}
}\\
\subfloat[]{
\includegraphics[width=\w]{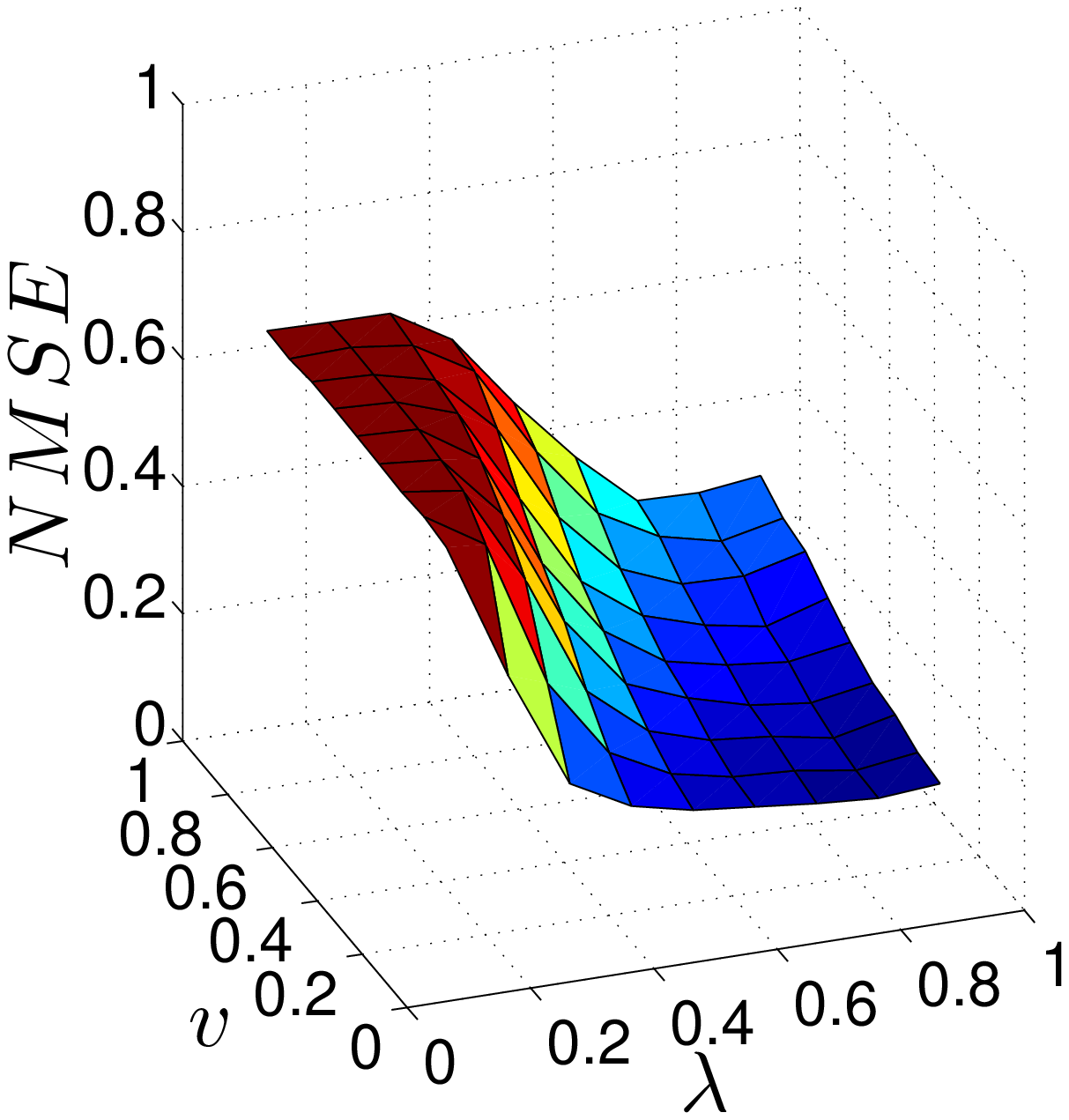}
\label{fig:esnnarmanonoise}
}
\subfloat[]{
\includegraphics[width=\w]{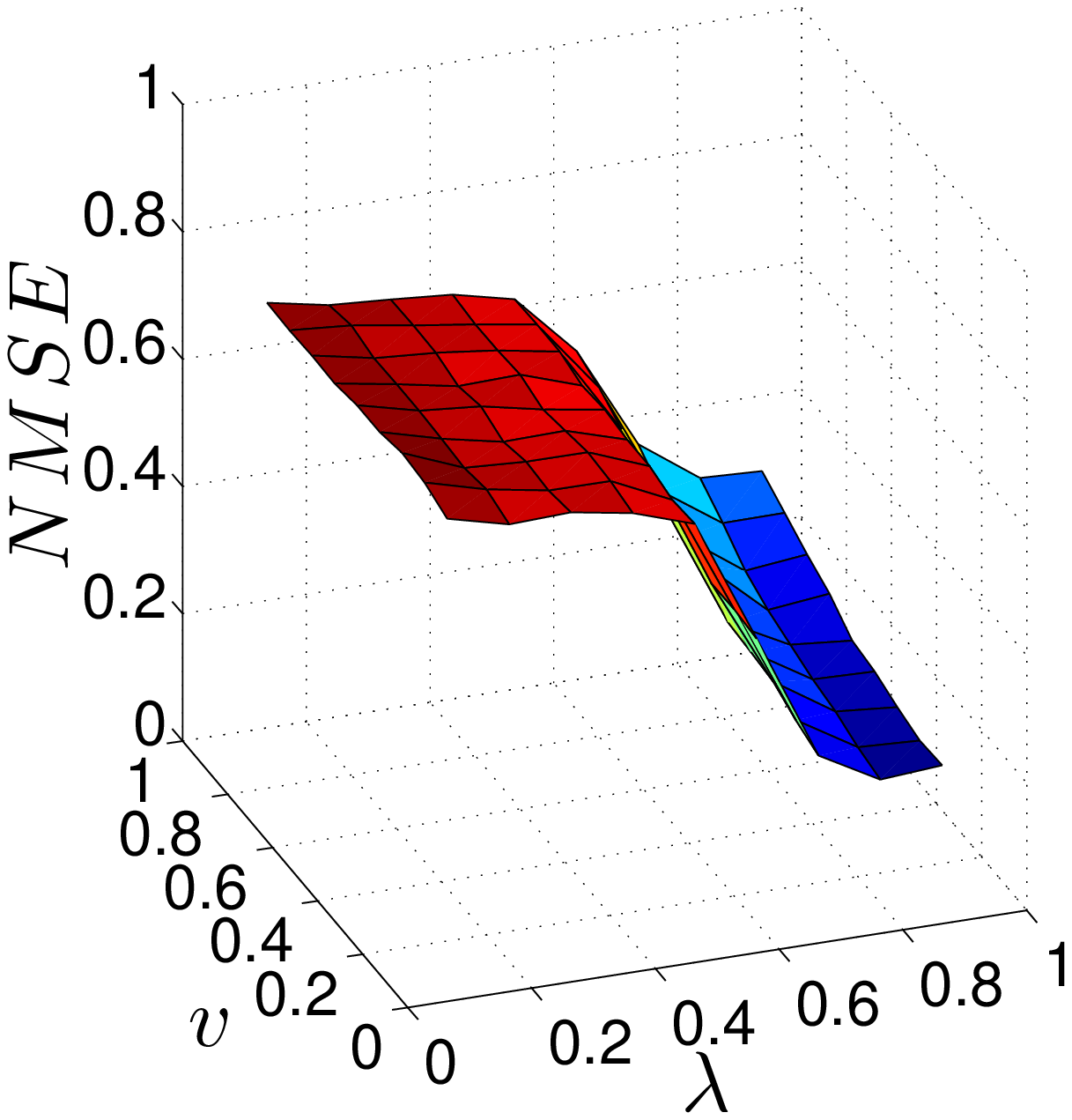}
\label{fig:esnnarmanoisy}
}
\subfloat[]{
\includegraphics[width=\w]{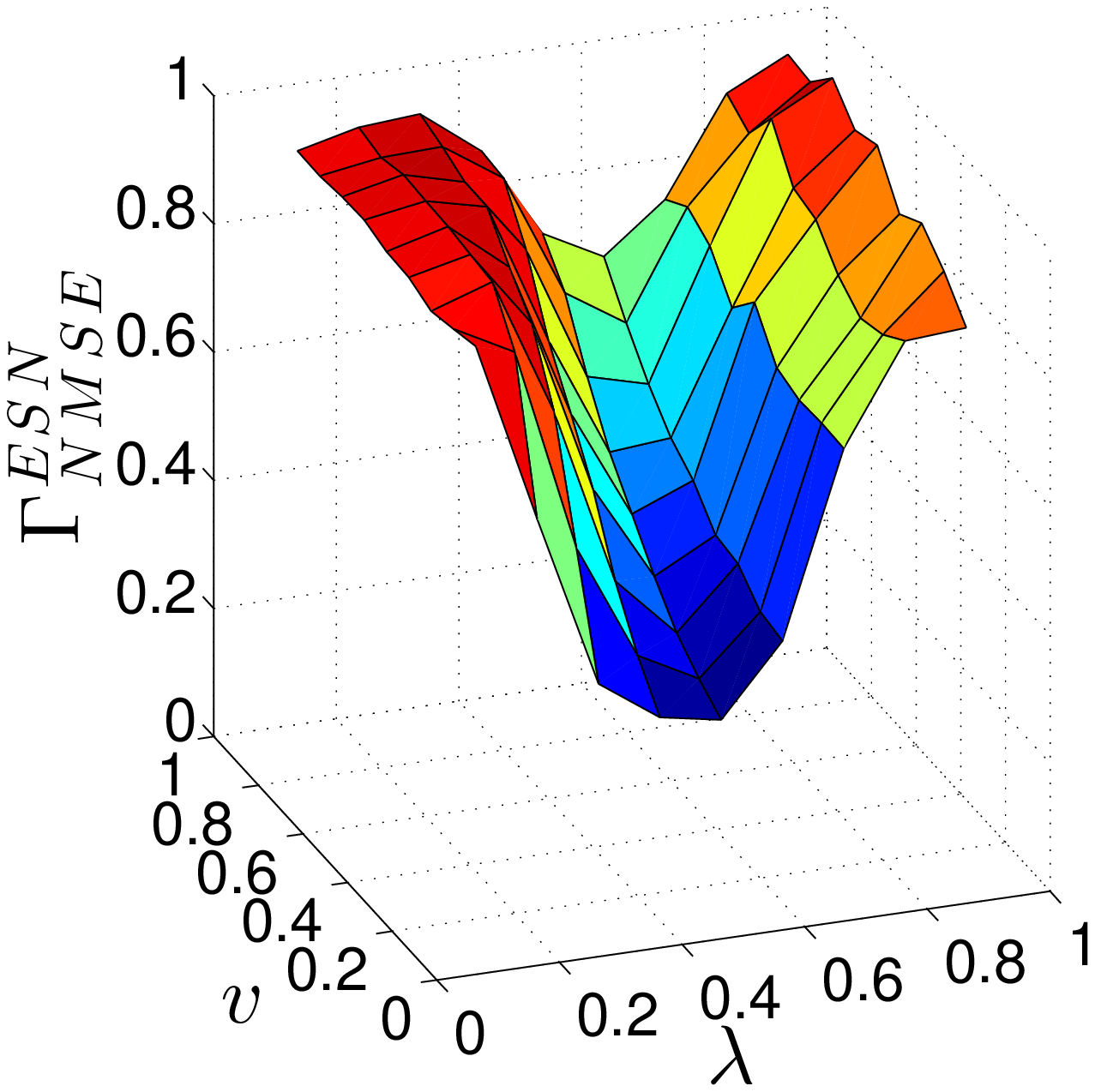}
\label{fig:esnnarmaratio}
}
\caption{Performance of ESN and SCR in solving the NARMA10 task measured using $NMSE$. The $NMSE$ in noise-free SCR (a) and SCR with structural noise (b). The ratio $\Gamma_{NMSE}^{SCR}(v,r)$ showing the overall variation of the performance between noisy and noise-free conditions (c). $NMSE$ in noise-free ESN (d) and ESN with structural noise (e). The ratio $\Gamma_{NMSE}^{ESN}(v,\lambda)$ showing the overall variation of the performance between noisy and noise-free conditions (f). For $r>0.8$, the SCR nonlinear task solving performance is completely robust to structural noise. ESN performance is also robust to noise for a critical spectral radius $\lambda>0.8$.}
\label{fig:mc}
\end{figure}

For the nonlinear computation NARMA10, we used SCRs  of size $N=100$ and plotted the testing error as a function of $v$ and $r$ (Figure~\ref{fig:scrnarmanonoise}). The best observed SCR performance ($NMSE=0.16$) occurs for  $r=0.9$ and $v=0.1$.  Figure~\ref{fig:scrnarmanoisy} shows the performance of noisy SCRs for which at every time step $n=2$ connections are perturbed with a white noise  with standard  deviation $\sigma=0.01$. For low $r$ and any $v$ the system performs poorly with $NMSE\approx0.8$. For $r>0.4$, there is a sharp drop in $NMSE$ and the system achieves an average optimal error of $NMSE=0.16$ for $r=0.9$ and $v=0.1$. Figure~\ref{fig:scrnarmaratio} shows the general effect of noise on SCR performance using the ratio $\Gamma_{NMES}^{SCR}$. We observe that for all $v=0.1$ and for $r=0.4$ this significantly reduces the performance to below 50\% of the original values while for $r>0.8$ the performance is virtually unaffected.
Figure~\ref{fig:esnnarmanonoise} shows the performance result of NARMA10 task for noise-free ESNs of size $N=100$ and connection fraction $l=0.2$. Similar to SCRs, the optimal performance is in the region $\lambda=0.9$ and $v=0.1$ with an average error of $NMSE=0.16$.  To test the performance of noisy ESNs when computing the NARMA10 task, $n=40$ reservoir connections are perturbed at each time step using identical white noise as before to achieve the same noise level.  Figure~\ref{fig:esnnarmanoisy} shows the result of this experiment. The optimal spectral radius  for noisy ESN does not change ($\lambda=0.9$ with average error $NMSE=0.19$). However, performance is very sensitive to spectral radius and for $\lambda <0.8$ shows a sharp increase in error. The effect of noise on the ESNs is summarized in Figure~\ref{fig:esnnarmaratio}. Networks with spectral radius $\lambda=0.5$  are affected the most and networks with $\lambda=0.9$ are robust. Compared with SCR, ESN is more sensitive to noise.  We can summarize the comparison between SCRs and ESNs using the following aggregate measures:
\begin{equation}
\widehat{NMSE}^{SCR} = \sum_{v\in V}\sum_{r \in R} NMSE(v,r) \text{,  }
\widehat{NMSE}^{ESN} = \sum_{v\in V}\sum_{\lambda \in \Lambda} NMSE(v,\lambda),
\end{equation}
and aggregate measures:
\begin{equation}
\widehat{\Gamma}_{NMSE}^{SCR} = \sum_{v\in V}\sum_{r \in R} \log(\Gamma_{NMSE}^{SCR}(v,r)) \text{,  }
\widehat{\Gamma}_{NMSE}^{ESN} = \sum_{v\in V}\sum_{\lambda \in \Lambda} \log(\Gamma_{NMSE}^{ESN}(v,\lambda)).
\end{equation}

For noise-free systems $\widehat{NMSE}^{SCR}=28.71$ and $\widehat{NMSE}^{ESN}=32.02$, showing that SCR outperforms ESN by $10.3\%$. For noisy systems, $\widehat{NMSE}^{SCR}=36.10$ and $\widehat{NMSE}^{ESN}=42.62$ which suggests the  simple structure of SCR makes it perform $15\%$ better than ESN in noisy environment. Finally, $\widehat{\Gamma}_{NMSE}^{SCR}=-17.40$ and $\widehat{\Gamma}_{NMSE}^{ESN}=-25.74$, indicating that SCRs are $1.47$ times more robust than ESNs for nonlinear task solving over the parameter space that we studied.

\begin{figure}[t]
\centering
\def \w {2.2in}
\subfloat[$N=100$]{
\includegraphics[width=\w]{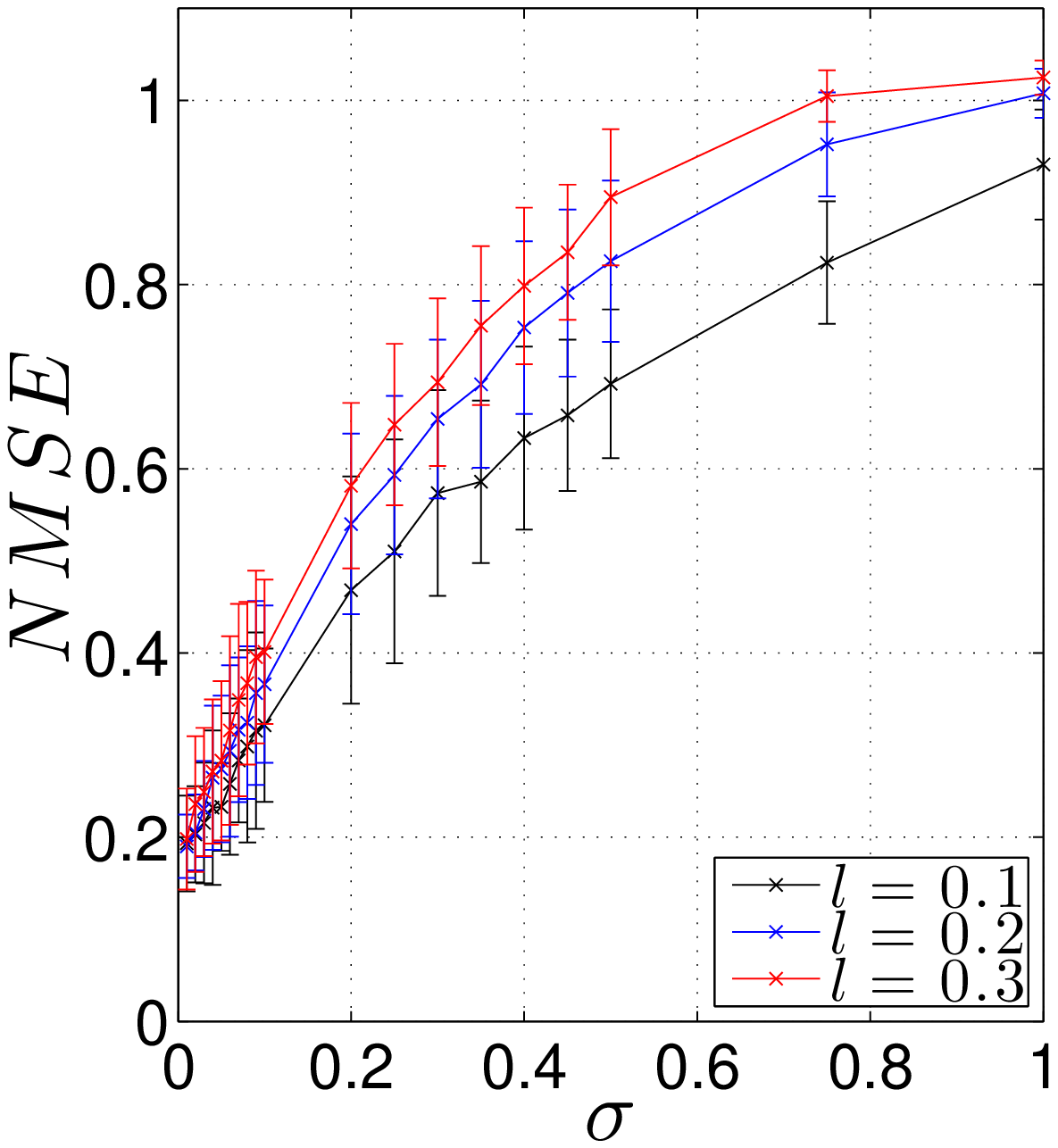}
\label{fig:varsp}
}
\subfloat[$l=0.1$]{
\includegraphics[width=\w]{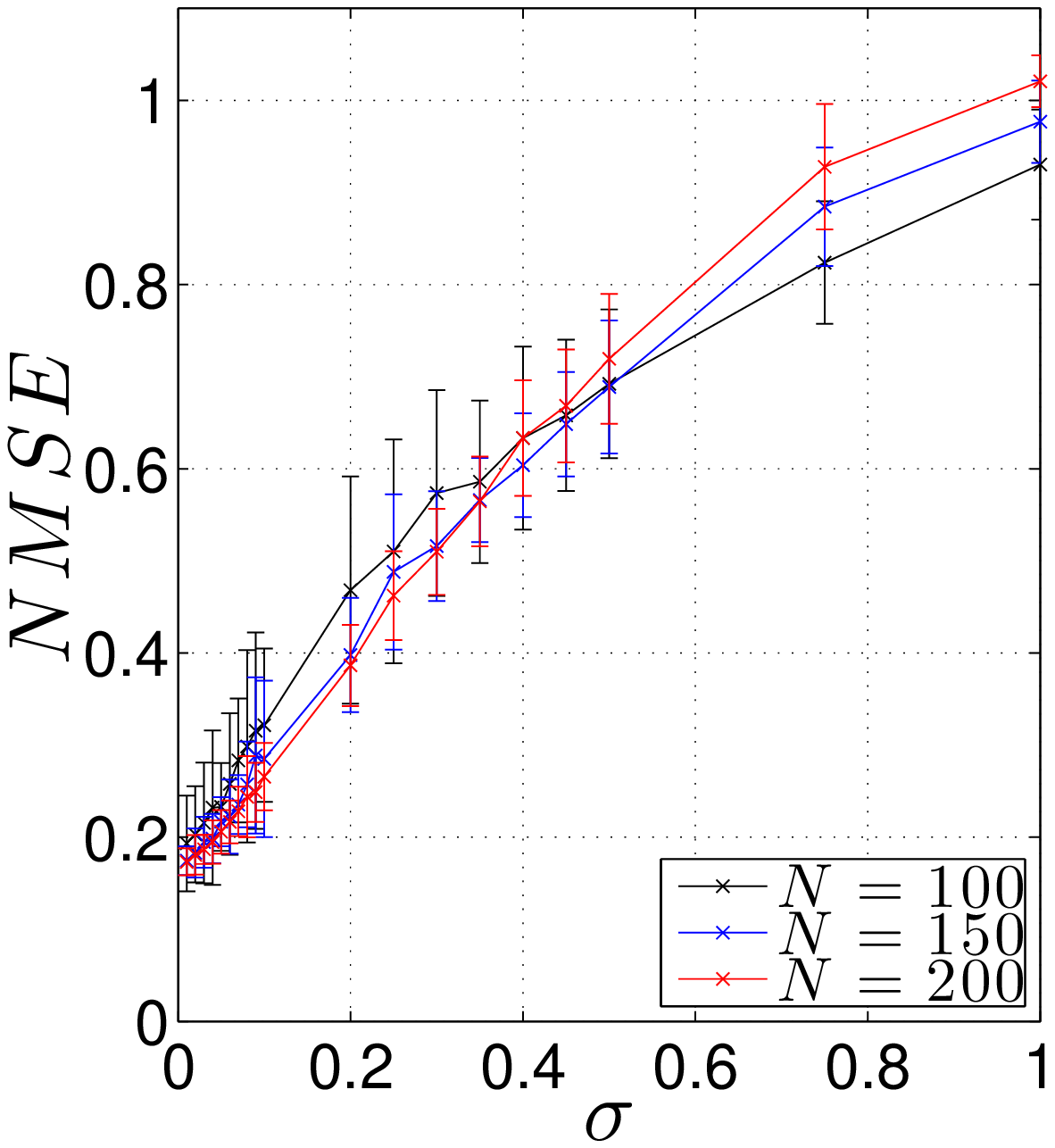}
\label{fig:varn}
}
\caption{Performance on the NARMA10 task as a function of standard deviation of noise $\sigma$. Under fixed  size $N=100$, networks with higher connection fraction $l$ lose their performance more quickly than sparser networks (a). Under fixed connection fraction $l=0.1$, the performance does not show significant sensitivity to network size as the noise level increases.}
\label{fig:sensitivity}
\end{figure}
Finally, we studied the sensitivity of ESN performance in the NARMA10 task under different noise levels $\sigma$, different network size $N$, and reservoir sparsity $l$ (Figure~\ref{fig:sensitivity}). For fixed network size, as we increase the connection fraction, the error increases more quickly as a function of noise (Figure~\ref{fig:varsp}). This is expected since in denser networks, variations in the state of one node  propagate to many downstream nodes. We hypothesize that if we control for node out-degree, we can contain this effect. We did not find any significant variation in the performance of networks with different sizes as a function of changing noise $\sigma$ (Figure~\ref{fig:varn}).

\section{Discussion}

We used theoretical models to investigate robustness of reservoir computing as an approach  to computation in emerging nanoscale and self-assembled devices. An example of such networks is  {\em Atomic switch networks} (ASN). These were based on a technology developed by Terabe et al.\cite{terabe2005} aimed at reducing the cost and energy consumption of electronic devices. They can achieve a memory density of 2.5\,Gbit\,cm$^{-2}$ without any optimization, and  a switching frequency of 1\,GHz. Recently, Sillin et al.\cite{0957-4484-24-38-384004} combined bottom-up self-assembly and top-down patterning to self-assemble ASN. These networks are formed using deposition of silver on pre-patterned copper seeds. They have a three-dimensional structure that contains cross-bar-like junctions, and can be transformed into metal-insulator-metal (MIM)  atomic switches in the presence of external bias voltage\cite{0957-4484-24-38-384004}. The morphology of this self-assembled network can be directed by the pitch and the size of the copper seeds, which control the density and wire lengths, respectively. We studied ESN and SCR with varying connection fraction, input weights, and spectral radius to model the controllable variables in ASNs. We also used a white noise to model variations in the electrical properties of nanowire networks   due to radiation or thermal noise. The normal distribution is known to be suitable to model variations in nanoscale devices\cite{4447311}. We showed that one can use the dynamical properties of a self-assembled system to perform computation without changing the microscopic structure of the system itself. The only modification to the structure of ESN and SCR is to adjust the spectral radius and therefore dynamical regime of the system, which is independent of the specific computation and can be done using external control signals\cite{doi:10.1162/neco.2007.19.1.111}.

\section{Conclusions}
We presented reservoir computing as an alternative approach to randomly assembled computers for implementing computation on top of  emerging nanoscale computing substrates. Using RC, we can compute with such devices assuming only enough connectivity in the system to propagate signals from the input to the output. This approach eliminates the need for control signals  and redundancy for programming and fault-tolerance in emerging architectures, which simplifies its implementation and makes the training more efficient. In addition, because the programming takes place in the output layer, the same device can be used to compute multiple functions simultaneously. We showed that the system resists noise in the interaction between nodes. This is a surprising feature because structural change in the system affects the long-term dynamics of the network. In RC with full input-output connectivity, the performance of SCR is similar to ESN. However, with sparse input-output connectivity the readout layer only has limited observation of the reservoir dynamics, therefore  the dynamics of different nodes in the reservoir have to be as independent as possible to represent independent spatiotemporal features of the input signal. In ESN, the reservoir nodes have more interactions and therefore their dynamics are more correlated resulting in a lower performance. In addition, with higher interactions between nodes, noise in a single connection can propagate to several other nodes, which distorts the dynamics of the ESN. In SCR, on the other hand, each node is only connected to one downstream node which limits the propagation of noise to only one other node. This result in higher robustness to noise in SCR. In future work, we will study this hypothesis by controlling the out-degree of ESN reservoir nodes.
 This is the first time RC has been used  to solve nonlinear tasks with sparse readout and structural noise. Exact characterization of performance and robustness under varying sparsity and weight distribution conditions is left for future work. Another future direction is implementation of a ``detect-and-recompute" schema as a fault-tolerance mechanism against one or more permanently failed nodes or connections.

\vspace{2mm}
\noindent{\bf Acknowledgments.}  This material is based upon work supported by the National Science Foundation under grant 1028238. M.R.L. gratefully acknowledges support from the New Mexico Cancer Nanoscience and Microsystems Training Center (NIH/NCI grant 5R25CA153825). We thank the anonymous reviewers for their constructive comments.

%
% ---- Bibliography ----
%

\bibliographystyle{splncs}
\bibliography{ucnc2014}

\end{document}